# Physical Layer Encryption for Industrial Ethernet in Gigabit Optical Links

Adrián Pérez-Resa, Miguel Garcia-Bosque, Carlos Sánchez-Azqueta and Santiago Celma

*Abstract*— Industrial Ethernet is a technology widely spread in factory floors and critical infrastructures where a high amount of data need to be collected and transported. Fiber optic networks at gigabit rates fit well with that type of environments where speed, system performance and reliability are critical. In this work a new encryption method for high speed optical communications suitable for such kind of networks is proposed. This new encryption method consists of a symmetric streaming encryption of the 8b/10b data flow at PCS (Physical Coding Sublayer) level. It is carried out thanks to an FPE (Format Preserving Encryption) blockcipher working in CTR (Counter) mode. The overall system has been simulated and implemented in an FPGA (Field Programmable Gate Array). Thanks to experimental results it can be concluded that it is possible to cipher traffic at this physical level in a secure way. In addition, no overhead is introduced during encryption, getting minimum latency and maximum throughput.

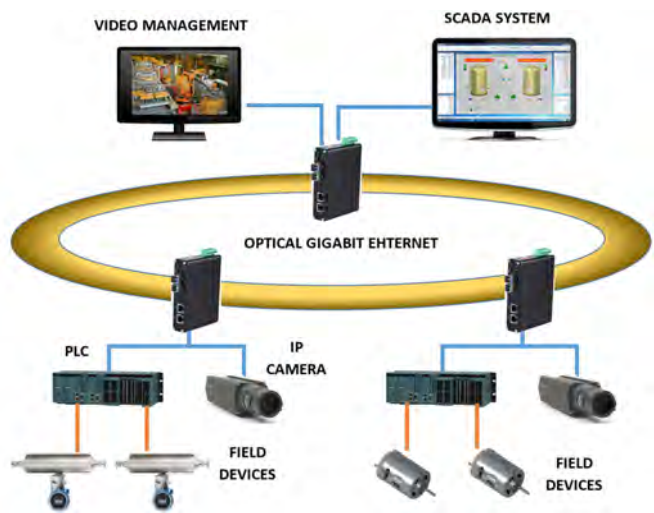

Fig. 1. Simple example of a SCADA network sharing field device traffic and video security.

## I. INTRODUCTION

THANKS to advances in Information Technology during the last decades, communication solutions at automation level in the ICS (Industrial Control Systems) have evolved from legacy field buses to Ethernet technology [1], [2], [3]. Ethernet has increased significantly its use as communication solution for SCADA (Supervisory Control And Data Acquisition) networks. On the other hand, not only data from industrial equipment is supported in SCADA networks, but also data traffic from surveillance video security and future IoT applications. Both kind of applications demand high transmission bandwidth. Therefore telecommunication equipment adapted for industrial environments with rates up to 1 Gbps and beyond are available from many vendors. A simple scheme of a SCADA network is shown in Fig. 1.

Manuscript received February 6, 2018; revised April 7, 2018 and May 13, 2018; accepted June 5, 2018. This work was supported in part by MINECO-FEDER under Grants TEC2014-52840 and TEC2017-85867-R and FPU fellowship program to M. Garcia-Bosque (FPU14/03523).

A. Pérez-Resa, M. Garcia-Bosque, C. Sánchez-Azqueta and S. Celma are with the Group of Electronic Design (GDE), Aragón Institute of Engineering Research (I3A), University of Zaragoza, Zaragoza, 50009, Spain (e-mail: aprz@unizar.es, mgbosque@unizar.es, csanaz@unizar.es, scelma@unizar.es).

Among different transmission media in Ethernet standards, optical fiber provides the best bandwidth, less signal loses, and more immunity to electromagnetic interference than other wired systems. Owing to this advantages many research related with multigigabit optical transceivers has been carried out for applications in industrial environments [4], [5].

In respect to wireless systems, optical communications can be considered safer as they do not emit any radiation that can be captured by an intruder. However, the optical signal can be intercepted easily thanks to fiber coupling devices and electro-optical converters [6]. Moreover it can be performed without perceptibly interfere in communications.

In layered communication models like OSI (Open Systems Interconnection) or TCP/IP, several encryption techniques can be used to avoid passive eavesdropping. It depends on the communication layer where confidentiality is needed. Many solutions proposed for industrial Ethernet encryption are usually for layers 3 or 4 such as IPsec or TLS (Transport Layer Security) standards, respectively. These offer security when field buses are attached to IP networks by means of a gateway [7]. Other solutions are proposed for layer 2, such as MACsec standard [8]. Moreover, industrial communication equipment, with layer 2 and layer 3 encryption capabilities are nowadays available from some vendors, and they are used to protect ICS from both internal and external cyber-attacks.

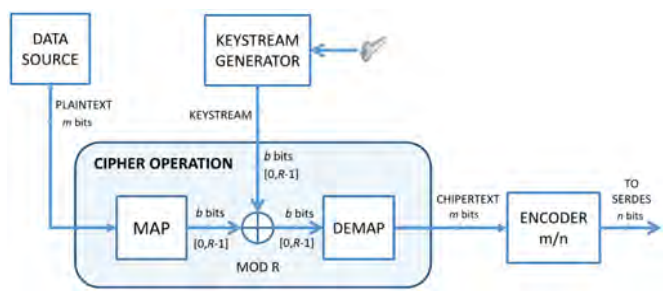

Fig. 2. Location and generic structure of a stream cipher in a physical layer with block line encoding.

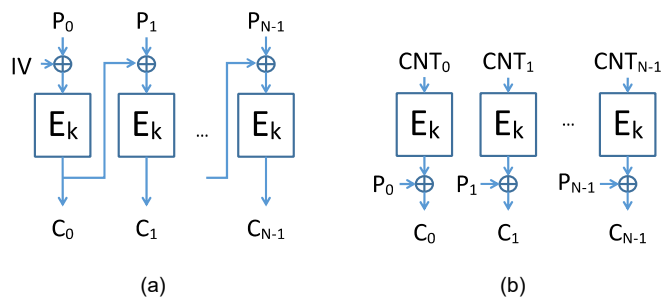

Fig. 3. Scheme of blockcipher operation modes (a) CBC, and (b) CTR. Plaintext is split in $N-1$ blocks of $B$ bits $\{P_0, P_1, …, P_{N-1}\}$ to get $N-1$ blocks for ciphertext $\{C_0, C_1, …, C_{N-1}\}$. IV is the initial value for CBC and CNT the counter for CTR mode, both are $B$-bit wide. $E_K$ is the underlying blockcipher and $B$ its blocksize in bits.

Regarding layer 1 encryption in optical communications, there is no proposal for Ethernet. However there are solutions related with optical technology such as OCDM (Optical Code-Division Multiplexing) [6], or with physical layer protocols, such as the encryption of OTN (Optical Transport Network) frame payloads [9]. The latter is mainly used by service providers to grant security on the carrier telecommunication networks.

Some advantages of performing encryption at layer 1 is that no extra data fields are introduced on ciphering packets, unlike protocols at other layers do [10]. Thanks to this, no throughput is wasted for transmitting this extra overhead. For example, OTN commercial equipment is usually able to perform encryption at line rate, achieving a 100% throughput and introducing very low latency [11], in the range of nanoseconds. It contrasts with other encryption methods such as IPsec which usually introduces latencies in the range of milliseconds.

Regarding to Ethernet, 1000Base-X is one of the most widely used physical layer standards at 1 Gbps rate in optical communications, and there is no mechanism providing layer 1 security on it. On the other hand, gigabit optical Ethernet is widely used in industrial Ethernet networks, where for certain applications a high level of determinism is needed, and there can be strict requirements for network delay and jitter [12]. A physical layer encryption mechanism could provide the mentioned advantages such as zero overhead and low latencies. This kind of 'in-flight' encryption could be useful with real time protocols, such as EtherCAT or PROFINET IRT [12], [13], enabling encryption to be performed at line rate.

The main motivation of this work is to propose and implement an encryption method suitable for the PCS (Physical Coding Sublayer) of the 1000Base-X standard. The physical coding of this Ethernet layer is the well-known 8b/10b encoding. To get 'in-flight' encryption of the 8b/10b symbol stream, a stream cipher that preserves the format of this codification is necessary.

As far as the authors are concerned, there is no standardized or recommended format preserving stream cipher. However, recently several FPE (Format Preserving Encryption) modes of operation have been approved by NIST (National Institute of Standards and Technology) [14]. The structures described in these modes can be understood as a kind of FPE "blockciphers" [15] that are able to encrypt data in the desired format. Its use in a like-of ECB (Electronic Code Book) mode is proposed to provide security in databases with an arbitrary format [15] or in legacy ICS protocols [16].

In this work, CTR (counter) mode is proposed to be used with "FPE blockciphers" to achieve security in high throughput applications where data format must be preserved. For example the Gigabit Ethernet 1000Base-X physical layer.

The paper is divided into the following sections. In Section II an introduction to PCS layer encryption necessities is given, in Section III an introduction to the FPE NIST recommendations is made. In Section IV security considerations of the proposed structure is given and the analysis of the keystream output is carried out. Subsequently, Section V deals with the practical case for Gigabit Ethernet 1000Base-X layer and the overall structure of the proposed encryption system. In section VI, the hardware implementation of the cipher is described and results of the encryption are explained. Finally, in section VII conclusions are given.

## II. CODING PRESERVING ENCRYPTION

Physical Coding Sublayer is part of the Ethernet 1000Base-X standard and performs functions such as autonegotiation, link establishment, clock rate adaptation and data encoding.

As other high speed standards, a baseband serial data transmission is carried out while clock frequency information is embedded in the serial bitstream itself. Thanks to bit transitions in the data stream the clock recovery circuits are able to extract the frequency information at the receiver.

Consequently, serial data sampling is made at the appropriate time. In order to facilitate the work of the CDR (Clock and Data Recovery) circuit, information must be encoded in such a way that a good transition density and a short run length are achieved. Also a DC-balanced serial data stream must be guaranteed by getting a good disparity.

Block line encoders group input bits into $m$-bit blocks and map these blocks into $n$-bit blocks where $n$ is greater than $m$. Thanks to the redundancy introduced in data, block encoders achieve the attributes mentioned previously [17] (transition

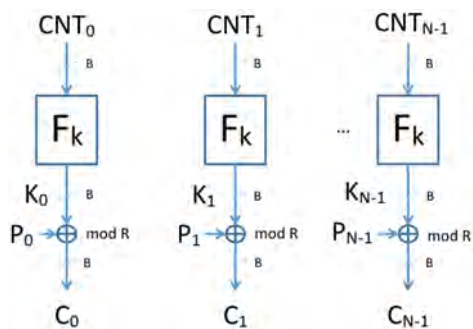

Fig. 4. Scheme of CTR mode using a FPE blockcipher $F_K$. Plaintext is split in $N-1$ blocks of $B$ symbols with radix $R$ {$P_0$, $P_1$, …, $P_{N-1}$} to get $N-1$ blocks for ciphertext {$C_0$, $C_1$, …, $C_{N-1}$}. Counter values {$CNT_0$, $CNT_1$, …, $CNT_{N-1}$} and keystream values {$K_0$, $K_1$, …, $K_{N-1}$} are also blocks of $B$ symbols with radix $R$. Modulo-$R$ addition is done symbol by symbol inside each block between keystrem and plaintext.

density, short run length and DC-balance). In the case of 1000Base-X, 8b/10b encoding is used.

Normally an additive stream cipher is implemented by carrying out the XOR operation between the plaintext and a keystream obtained from a secure pseudorandom generator. In case of encrypting physical layer when block-line encoding is used, it is necessary to preserve the mentioned encoder properties, therefore the location of the encryptor in the datapath must be taken into account. In the case of a block line code such as 8b/10b, encryption should be carried out before the encoder and the XOR operation would not be suitable. An $m$-bit symbol should be encrypted giving rise to another $m$-bit symbol, within the alphabet of symbols supported by the encoding standard. If $R$ is the number of possible symbols, then the operation performed by the stream cipher should be a modulo-$R$ addition instead of an XOR. Moreover $m$-bit symbols should be mapped in a value between $0$ and $R$-$1$ before this addition and the resulting value reverse-mapped to the corresponding new $m$-bit symbols that are finally encoded.

In Fig. 2 the generic mechanism is shown with such kind of encoding. The $m$-bit symbols are mapped to $b$-bit values that are encrypted, reverse-mapped and finally encoded into $n$-bit words. In addition to these operations, the keystream generator must be able to produce a uniformly distributed keystream in the range of the available alphabet of symbols. To achieve this, in this work CTR mode is proposed to be used with FPE blockciphers. FPE modes are introduced in Section III, and security considerations of the overall structure are given in Section IV.

### III. FPE CIPHER MODES

Format Preserving Encryption encrypts plaintext in a ciphertext preserving its original format and length. Some examples where this type of encryption is useful are the Primary Account Numbers (PANs) or Social Security Numbers (SSNs) for which standard block ciphers or their operating modes would not preserve the format. Currently two operation modes for FPE are recommended by NIST, FF1 and FF3 [14]. Both "ciphers" (actually they are considered operation modes) have a scheme based on a non-binary Feistel structure. In NIST recommendations the underlying round function of the Feistel network consists of an AES (Advanced Encryption Standard) blockcipher, although in its original specifications this PRF (Pseudo Random Function) can be also implemented by a hash function based on HMAC (Hash-based Message Authentication Code) [18].

The main differences between both modes, FF1 and FF3, are the number of rounds in the Feistel structure (with a smaller number in FF3); and the way to achieve the arbitrary length message encryption. In the case of FF1, its specification allows message lengths up to $2^{32}$ words, while in FF3 message length is more limited and it is constrained by the input data radix. However, in the original proposal for FF3, also called BPS (Brier Peyrin Stern), the basic blockcipher component BPS-BC is proposed to be used in CBC (Cipher Block Chaining) mode for encryption of arbitrary length messages [14]. In spite of this, CBC operation mode properties result in frequent misuse among classical confidentiality block cipher modes. Indeed, CTR can be considered the best and most modern way to achieve privacy-only encryption [19]. The basic operation for both modes is shown in Fig. 3.

As far as the authors know there is no stream cipher proposal able to preserve the format of the plaintext symbols. The main stream cipher proposals are oriented to generate binary keystreams, as the finalists of eSTREAM project [20] or the well-known solution formed by secure blockciphers in CTR mode. Due to this absence, we consider that the use of the recent FPE modes approved by NIST in conjunction with CTR mode could be a good solution.

### IV. SECURITY CONSIDERATIONS

#### A. FPE with CTR Mode

Assuming that the plaintext is a stream of symbols defined with radix $R$, we can define the CTR mode for our purpose as described by NIST [21] but using a modulo-$R$ addition as encryption operation instead of an XOR. For a family of functions $F$ such that:

$F: K \times \{0,1,\dots,R-1\}^B \to \{0,1,\dots,R-1\}^B$ where $B$ is the block size and $K$ the keyspace, and given a plaintext $P = \{P_0, P_1, \dots, P_{l-1}\}$ with $l > B$, $P$ is divided into $N$ blocks $P_i$: $N$-1 blocks with $B$ symbols plus one block with the rest. The $N$ blocks of the ciphertext $C_i$ are obtained applying Algorithm 1.

| Algorithm 1. CTR mode for FPE encryption |
|---|
| $CNT_0 = INIT\_CNT$ |
| $K_i = F_K[CNT_i]\ for\ i = 0,1,\dots,N-1$ |
| $CNT_{i+1} = CNT_i + 1\ for\ i = 0,1,\dots,N-1$ |
| $C_i = (P_i + K_i) mod\ R\ for\ i = 0,1,\dots,N-2$ |
| $C_{N-1} = (P_{N-1} + MSB(K_{N-1})) mod\ R$ |

where $N$ $K_i$ blocks are the keystream obtained from $N$ successively counter values $CNT_i$. $MSB(K_{N-1})$ are the Most Significant Symbols of the last keystream block that operate with the remaining symbols in the last plaintext block $P_{N-1}$. The modulo-$R$ addition is performed symbol by symbol between each plaintext and keystream blocks. In Fig. 4 the proposed scheme is shown.

The decryption algorithm will be as Algorithm 1, but replacing the plaintext $P$ with the ciphertext $C$, and the modulo-$R$ addition by a modulo-$R$ subtraction.

According to [22] and posterior studies it can be shown that in the sense of IND-CPA (indistinguishability against chosen-plaintext attacks) security, given an adversary $X$ attacking a CTR scheme there exists another adversary $Y$ attacking the PRF security of the underlying function $F_K$ of that CTR scheme. The advantages, $ADV$, of such adversaries are related as shown in (1).

$$ADV_{CTR}^{IND-CPA}(X) \leq 2 \cdot ADV_F^{PRF}(Y) \quad (1)$$

Also according to its proof, it can be concluded that it holds independently of the radix $R$ on which the CTR mode operates. On the other hand, according to the PRF-PRP switching lemma [22] and since the underlying function $F_K$ is a blockcipher that can be considered a PRP, it is well-known that the cited advantage is degraded by a term of $Q^2/2^{B+1}$, as shown in (2).

$$ADV_{CTR}^{IND-CPA}(X) \leq 2 \cdot (ADV_F^{PRP}(Y) + Q^2/2^{B+1}) \quad (2)$$

where $Q$ is the number of queries made by the adversary $Y$ and $B$ the blocksize of the blockcipher. In the case of an FPE blockcipher with radix $R$ this factor will be $Q^2/2R^B$. Indeed, due to the birthday paradox, it could be considered insecure to encrypt more than $Q = \sqrt{R^B}$ blocks with the same key.

It is possible to conclude that, in the same way that the use of CTR mode with a standard blockcipher is considered safe, we can safely use the CTR mode with an FPE blockcipher. On the other hand, it will be important to take into account that the block size $B$ is a configurable parameter for FPE blockciphers, and it affects the overall security of the CTR scheme.

One requisite for $B$ could be provide the same security limit as a recommended blockcipher with a 128-bit blocksize. For AES working in CTR mode $Q = 2^{64}$, then the data limit to be transmitted before key refreshing should be lower than $L = Q \cdot 128 = 2^{71}$ bits. As the proposed system in this work should have at least the same bound we have taken as design constraint the requisite shown in (3).

$$L = Q \cdot B \cdot b \geq 2^{71} \rightarrow \sqrt{R^B} \cdot B \geq \frac{2^{71}}{b} \quad (3)$$

where $B$ is the blocksize in symbols defined with radix $R$, and $b$ the information bits represented by each symbol in the encoding standard. This constraint gives a lower bound for $B$

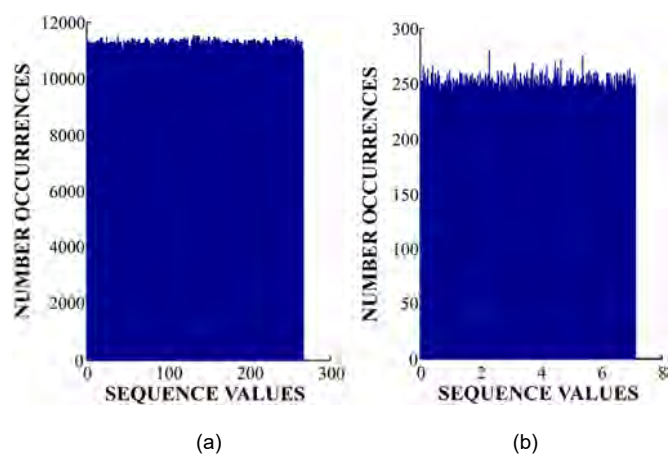

Fig. 5. Histograms obtained in: (a) frequency test and (b) serial test. Sequence values units in (b) are given in thousands.

and it is taken into account in Section V where the cipher parameters are explained.

### B. Keystream Analysis

It is well known that one of the security criteria of a stream cipher is based on the fact that the generated keystream is indistinguishable from a truly random sequence. Therefore it is useful to check the randomness of the obtained keystream sequence $\{K_0, K_1, \ldots, K_{N-1}\}$ in Fig. 4.

Among the most used tests for randomness, we can highlight some of them such as NIST [23], Diehard [24] or TestU01 [25]. All of them need as input a binary stream or a sequence of integers within a range that is a power of two. In our case the keystream values are not included in a range like that due to the encoding used. The range of possible values for the keystream will be explained later in Section V and VI. As far as the authors know, there is no standardized set of tests able to check the randomness of a sequence of integer numbers that are not included in a power of two range. However, the battery of tests proposed by Knuth [26] does include different tests that are suitable for our purpose, as they do not have the mentioned constraint.

In this work the keystream resulting from our FPE implementation was evaluated applying the following tests described in [26]: frequency test, serial test for pairs and triples of symbols, poker test, run and serial correlation test.

Regarding to frequency and serial test, Chi-Square Goodness of Fit test was applied successfully with 5% and 1% of significance level. Histograms for the keystream sequence organized in tuples of one and two symbols are shown in Fig. 5a and Fig. 5b for sequences of 3 and 15 millions of tuples, respectively.

As for Poker test, also Chi-Square Goodness of Fit test was applied using the five categories described by Knuth [26].

With respect to run test, it was applied to runs up and down successfully for sequences of 5000 values.

Correlation coefficient was calculated for a window of 25000 samples and shifts up to 100000, giving a result

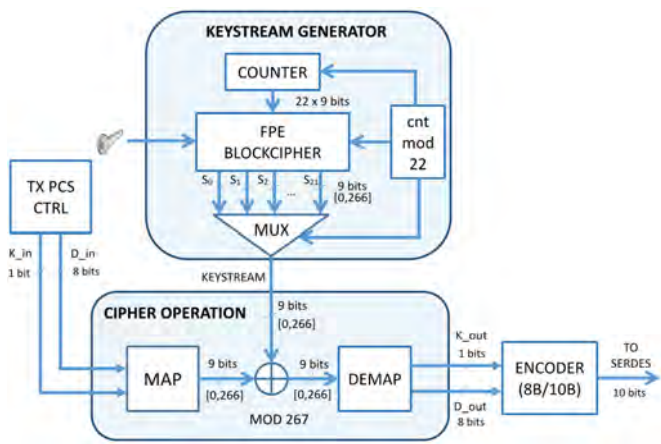

Fig. 6. Overall structure for the streaming encryption system in a physical layer with 8b/10b encoding. Decryption will be as encryption but using a modulo-267 subtraction instead of an addition.

between the bounds recommended for this test.

Although the fulfillment of a statistical test does not guarantees that the sequence is truly random, we can conclude that the generated sequence is suitable as keystream for the proposed stream cipher.

## V. APPLICATION CASE: ETHERNET 1000BASE-X

In this paper we have focused on the case of high-speed communications, particularly in the 1000Base-X standard used in optical Gigabit Ethernet links. In this standard, the PCS (Physical Coding Sublayer) level is responsible for generating and encoding control and data symbols that are transmitted to the optical line. As the block line coding is 8b/10b the purpose of the encryption will be to cipher the complete 8b/10b symbol flow as shown in Fig. 2.

The main advantage of the proposed method is that there is no throughput loss because no extra data fields are added to the packets when are encrypted, then no overhead is introduced. Moreover it not only encrypts the contents of the packets but also the activity or data traffic pattern. This is because by encrypting at 8b/10b symbol level, control symbols and ordered sets are also encrypted, such as packet start and end symbols or IDLE sets in the IFG (Inter Frame Gap). This last capacity could improve security as it prevents passive eavesdroppers from performing traffic analysis attacks. It would be useful in scenarios where traffic pattern analysis could reveal sensitive information about the behavior of a critical infrastructure or facility.

The data set to be encrypted is a limited set, since the valid 8b/10b symbols are composed of 256 data symbols plus 12 control symbols, a total of 268 symbols which do not generate code errors. On the other hand, the special control symbol /K28.7/ used only in diagnostic functions is capable of generating an undesired Ethernet "comma" sequence if it coincides sequentially with some particular symbols [27]. A comma across any two adjacent code-groups may cause code-group realignment. Symbol /K28.7/ is not used for standard data communication, and in order to avoid the accidental generation of it in the encryption of any 8b/10b symbol it has been excluded from the encryption symbol mapping.

Therefore, the valid set of symbols consist of 267 possible values to be encrypted. As in Fig. 2, in order to carry out the encryption, the 267 possible symbols are mapped in the range of 0 to 266, giving a value of 267 for radix parameter $R$. After symbol mapping, modulo-$R$ addition is performed with the keystream. Once the cipher operation is done, the resulting values are reverse-mapped to the corresponding new 8b/10b symbol. The new symbols will generate neither code error nor realignment as they exist in the set of 267 possible symbols and /K28.7/ has been excluded from it.

To generate the keystream, an FPE blockcipher in CTR mode has been built. Among the two NIST recommendations, FF1 and FF3, we have selected the latter. FF1 mode is not as limited in block length as FF3, since its specification says that it can reach up to $2^{32}$ symbols, then a greater security could be achieved. However, since for FF3 mode fewer rounds are required in the Feistel network, the cost in hardware resources is lower. In this mode the block size is limited according to the radix used, as shown in (4).

$$\begin{array}{c} R \in [2 \ldots 2^{16}] \\ R^{minlen} \geq 100 \\ 2 \leq minlen \leq maxlen \leq 2 \lfloor \log_R(2^{96}) \rfloor \end{array} \quad (4)$$

where $R$ is the radix and $minlen$ and $maxlen$ the bounds for the block size $B$. Since radix is 267, according to this equation block size $B$ is between 2 and 22.

On the other hand, according to (1), for $R = 267$ and $b = 8$ information bits/symbol, $B$ should be greater than or equal to 16, therefore it is possible to achieve this value using FF3 mode, as 22 is the maximum allowable value for $B$.

In this work the selected value for block size is the maximum, it means 22. The maximum block size means a greater number of cycles available per stage of a possible pipelined architecture, which allows a better reuse of the resources consumed by the hardware.

Given these parameters, the structure of the full streaming encryption system is shown in Fig. 6. It is similar to Fig. 2, but the keystream generator module has been replaced with the final CTR structure based on FPE blockcipher.

In Fig. 6. TX_PCS_CTRL module represents the PCS (Physical Coding Sublayer) controller that generates the 8b/10b symbol flow. Each symbol is formed by its 8-bit value $D\_in$, and the control flag $K\_in$. This control flag signals if the 8b/10b symbol is a control or data one, depending if its value is '1' or '0', respectively. Both signals, $D\_in$ and $K\_in$ are the input to the cipher. The encrypted output is a new 8b/10b symbol formed by $D\_out$ and $K\_out$, and it is the input of the encoder.

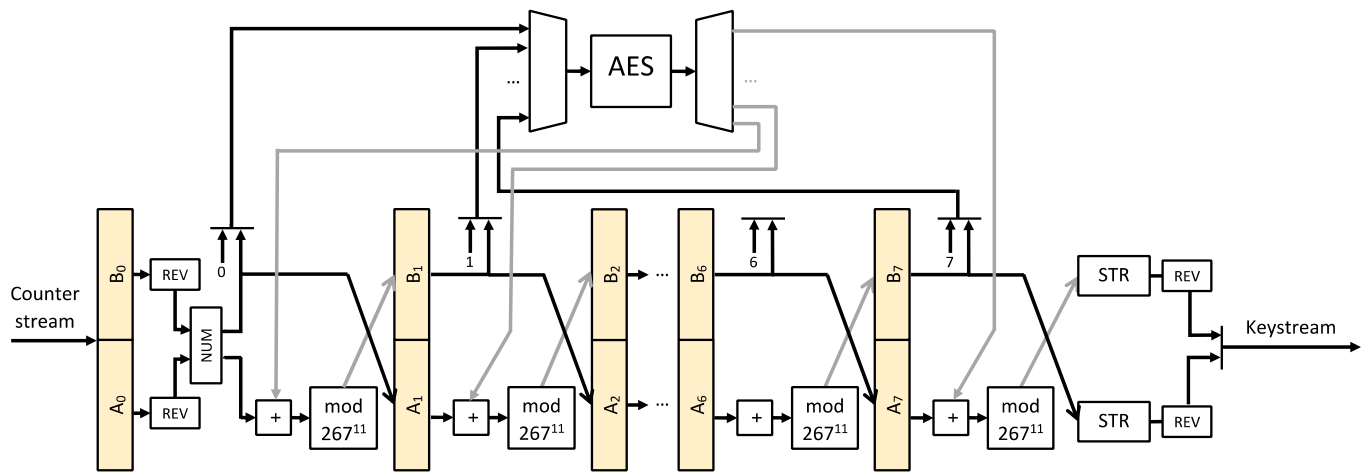

Fig. 7. Hardware structure of the implemented FPE blockcipher of Fig. 6. Its input is the stream of counter values and its output the keystream.

TABLE I: COMPARISON WITH OTHER SOLUTIONS

| | FF1 [16] | FF3 [16] | This Work |
|---|---|---|---|
| Slice Registers | 11285 | 5592 | 11127 |
| Slice LUTs | 7426 | 3587 | 16978 |
| 18K Block RAMs[1] | 343 | 170 | 77 |
| Slices[2] | 3268 | 1596 | 5636 |
| Operation. Freq. (MHz) | 279.6 | 283.5 | 125 |
| Cycles/Encryption | 707 | 269 | 1 |
| Bytes/Encryption | 13 | 13 | 1 |
| Encryption Rate (Mbps) | 41.1 | 109.6 | 1000 |
| Encryption Rate/Slice (Kbps/Slice) | 12.57 | 68.7 | 177.4 |

[1]77 Block RAMs are used in this work is due to the AES core. [2]Slices are estimated from the number of register and LUTs, assuming they are not packed together.

In this structure, for each clock cycle a modulo-22 counter (called 'cnt_mod_22' in Fig. 6) selects one of the 22 symbols from the output block of the FPE blockcipher. In Fig. 6 these symbols are represented by the set $\{S_0, S_1, ..., S_{21}\}$. The selected output symbol will be added to the incoming plaintext thanks to the modulo-267 addition. The FPE output block is refreshed every 22 cycles. The same happens with its input block that comes from CTR counter (called COUNTER in Fig. 6). This COUNTER also will increment its value every 22 cycles.

## VI. SYSTEM IMPLEMENTATION

### A. System Description

FF3 encryption algorithm is described in [14] and it implements a non-binary Feistel network. In this work, the cipher tweak value has been set to zero, and only the key is configurable. By taking into account the parameters that we need for our system (radix $R = 267$ and block size $B = 22$) the hardware structure for the FPE blockcipher can be particularized and represented as shown in Fig. 7, where the main functions are implemented in modules REV, NUM, STR, AES and $mod\ 267^{11}$, all of them described in [14]. In this implementation AES module has a key size of 128 bits.

The width of the data bus in each branch of the Feistel network is 11 symbols, half of the block size.

This design has a pipelined architecture to achieve the required throughput, as blocksize is 22, every 22 cycles 22 symbols are generated in the blockcipher output and each stage in the pipeline can last at most 22 cycles. Each round can be implemented in several stages and its number depends on the functions that it carries out. Particularly, in this implementation, function REV takes zero stages, NUM and AES one stage, STR 10 stages, and $mod\ 267^{11}$ two stages.

### B. Implementation Results

The system described in Fig. 7 has been implemented in a Xilinx Virtex 7 FPGA. Regarding the resources used for the FPE blockcipher, these are shown in Table I in terms of LUTs (Look-Up Tables), registers, and Block RAMs. In this design, no DSP (Digital Signal Processing) cell has been used.

Table I also shows a resource comparison with other implementations, particularly with [16], where FF1 and FF3 modes are implemented. Such implementations use an iterative looping architecture that saves resources but increments the number of cycles to carry out one block encryption. Although it is difficult to establish a comparison, it has been made in terms of throughput/resources. Among the three modes studied in [16], only FF1 and FF3 mode have been selected for the comparison, as FF2 was removed from NIST recommendation.

As shown in Table I the proposed system in this work entails a ratio throughput/resources better than existing FPE implementations. This comparison is only orientative as FPGA devices used in both implementations are different. While in [16] implementation is made over a Virtex-6 device, in this work Virtex-7 is used. Anyway, in both devices CLB structure is similar in terms of LUTs and registers, with four 6-input LUTs and 8 registers per slice.

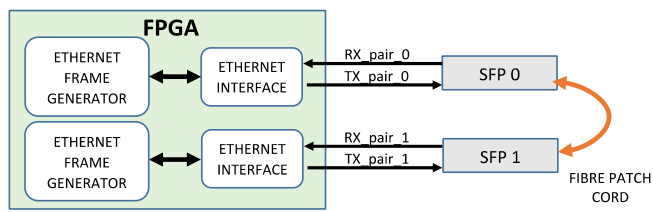

Fig. 8. Test set-up scheme.

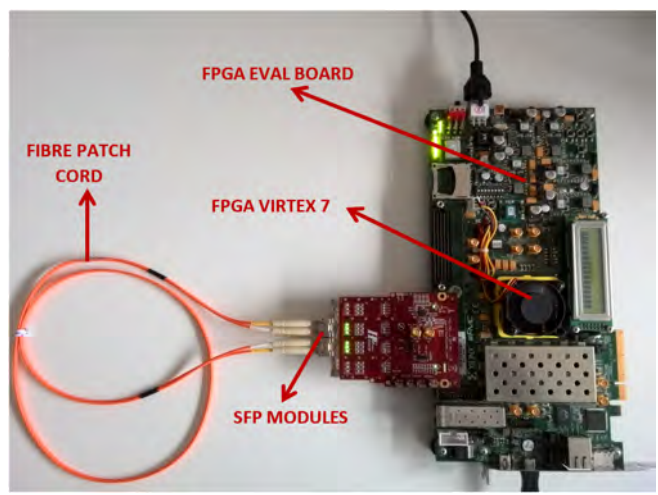

Fig. 9. Test set-up photo with SFP modules working at 1 Gbps.

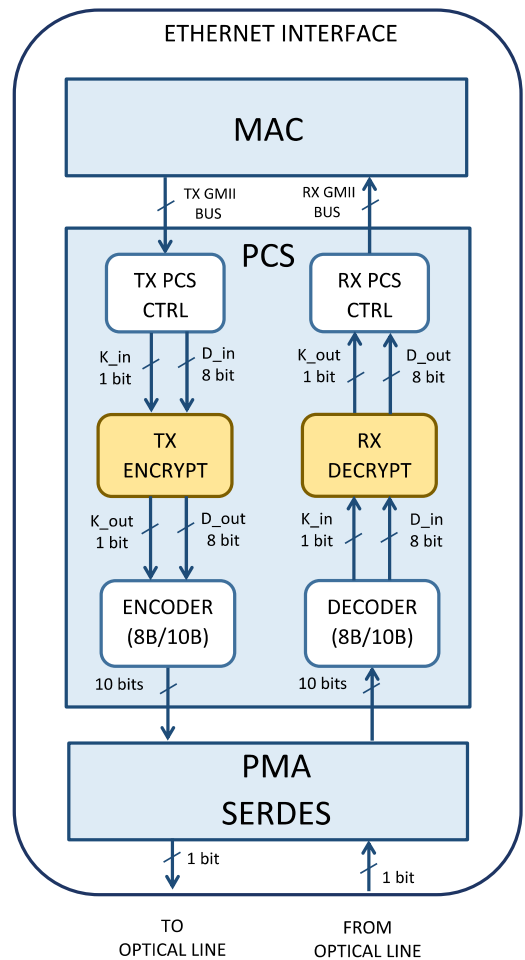

Fig. 10. Structure of the Ethernet Interface. It is composed by the MAC module, the PCS and the SERDES. In the PCS layer TX_ENCRYPT and RX_DECRYPT are the encryption/decryption modules. Both include the CIPHER_OPERATION and KEYSTREAM_GENERATOR modules shown in Fig. 6. TX_PCS_CTRL and RX_PCS_CTRL are the PCS controllers in charge of generating and receiving the 8b/10b symbols.

Also, it is important to remark that in this work the necessary radix is 267. It implies that $mod\ 267^{11}$ and STR modules are more complex to implement than the cases where radix is a power of two. In the latter division and modulo operations could simply be reduced to shifting and slicing operations over bit vectors.

### C. Encryption Set-Up

To test the proposed encryption system, the overall system described in Fig. 6 has been integrated in a 1000Base-X Ethernet interface linked to an Ethernet frame generator. Two chains composed each one by an Ethernet interface and Ethernet frame generator have been implemented over a Xilinx Virtex 7 FPGA. In the set-up for test the Ethernet interfaces have been connected to two SFP (Small Form-Factor Pluggable) modules suitable for Gigabit Ethernet standard over multimode fiber. The overall set-up scheme and a photo of the hardware system are shown in Fig. 8 and Fig. 9, respectively. In Fig. 10 the structure of the Ethernet interface is shown. Ethernet frame generators have been used to check the encrypted link with real traffic and verify that no frames are lost and no CRC (Cyclic Redundancy Check) errors are produced during encryption.

The final PCS structure, shown in Fig. 10 contains the encryption and decryption modules. In the 8b/10b TX datapath the latency introduced by the encryption module TX_ENCRYPT is 192 ns, and approximately the same for the RX datapath. In addition, no overhead is introduced in the encryption process, then a 100% throughput is achieved.

These values of latency and throughput are a doubtless improvement with respect to other encryption mechanisms, such as IPsec. As an example, in [28] latency measurements made over commercial equipment that implements IPsec give values between 50 and 350 ms. Moreover the inherent overhead introduced by IPsec during encryption limits the throughput of these equipment to values between 20% and 90% of the maximum achievable.

### D. Encryption Results

Different Ethernet traffic patterns have been tested. Particularly, no frame transmission and transmission of frames with randomized payload at different rates have been encrypted. Patterns have been named A, B, C and D. A corresponds with the case of no frame transmission, where only IDLES are transmitted over the link. B, C and D correspond to continuous frame transmission of 1024-bytes

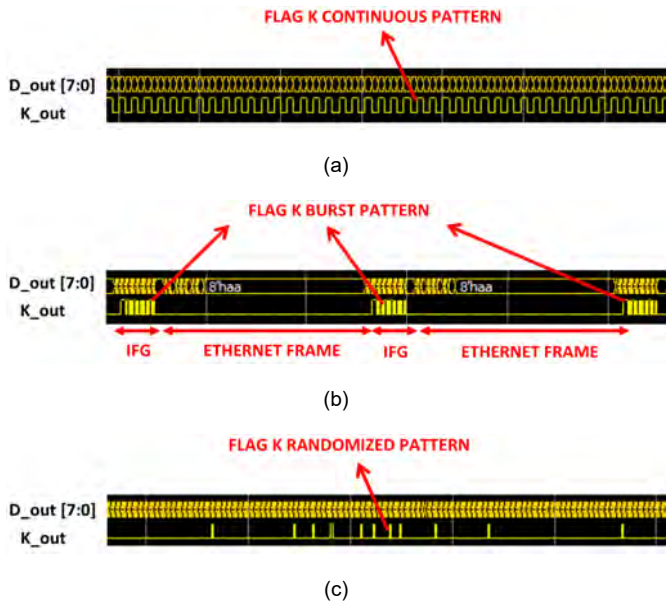

(a)

(b)

(c)

Fig. 11. (a) K flag pattern without encryption when no Ethernet frame is transmitted; (b) K flag pattern without encryption when transmitting an Ethernet frame burst; (c) K flag pattern after encryption regardless of the transmission or non-transmission of Ethernet frames.

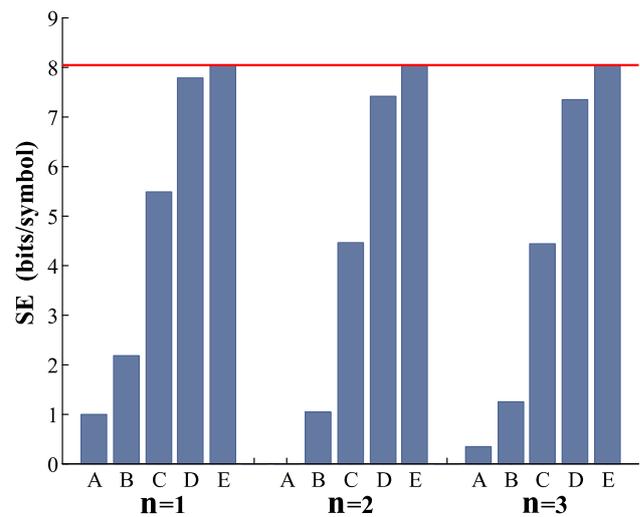

Fig. 12. Shannon Entropy measured with n from 1 to 3, and for all mentioned Ethernet traffic patterns, A, B, C, D and E.

length at rates of 10.2%, 50% and 91% of the maximum Gigabit line rate, respectively.

As mentioned previously, the proposed encryption system is able to make indistinguishable data traffic patterns. For this capability is interesting to monitor the signal waveform at the input of the 8b/10b encoder. As shown in Fig. 6 and Fig. 10, the output of the encryption is the input of the encoder, and it is formed by the 8b/10b symbol value $D\_out$ and its K control flag (named $K\_out$). Each 8b/10b symbol is a control or data one depending wheter its K flag is '1' or '0', respectively. $D\_out$ and $K\_out$ are used by the 8b/10b encoder to encode an 8-bit symbol into a 10-bit one.

K control flag pattern has been monitored for each of the mentioned traffic patterns (from A to D) with the encryption activated and deactivated.

When the encryption is not enabled in the case of A pattern, where no frames are transmitted, K control flag waveform is a signal that switches continuously between '0' and '1', as shown in Fig. 11a. It is because when no frame is transmitted, the physical layer controller (TX_PCS_CTRL module in Fig. 10) transmits continuously IDLE sets to the 8b/10b encoder. These IDLE sets are composed by two consecutive symbols, the control symbol /K28.5/ (with K flag equals to '1') plus a data symbol (with K flag equals to '0').

However, with encryption disabled for patterns B, C and D, K flag waveform seems a blast signal, as IDLE transmission only occurs in the IFG between frames, and during frame transmission only 8b/10b data symbols are transmitted (setting K flag continuously to '0'). It is shown in Fig. 11b.

On enabling encryption, K flag waveform is completely randomized in all cases, A, B, C and D, making indistinguishable which pattern is being transmitted, as shown in Fig. 11c.

As an overall result of encryption, SE (Shannon Entropy) has been measured for each of the mentioned patterns before and after encryption, as defined in (5).

$$SE = -\frac{1}{n} \cdot \sum_{\beta_n \in R^n} P(\beta_n) \cdot \log_2 P(\beta_n) \quad (5)$$

The 8b/10b symbol stream for each traffic pattern, mapped between 0 and $R$-1, has been grouped in tuples of $n$ symbols, called $\beta_n$, and the probability for each tuple, $P(\beta_n)$, has been calculated. Particularly, SE has been measured for values of $n$ from 1 to 3 in each of the mentioned patterns A, B, C and D. For each $n$, 1, 2 and 3, the length of the used sequences has been 2.67, 14.26 and 571 Msamples, respectively.

Because in (5) logarithms with base 2 are used, SE result is given in bits/symbol. Ideally, if every $n$-tuple ($\beta_n$) is equally likely with probability $P(\beta_n) = p = R^{-n}$ the value of Shannon Entropy for every $n$ should be as in (6).

$$-\frac{1}{n} \cdot R^n \cdot p \cdot \log_2 p = \log_2 R = \log_2 267 \cong 8.0606 \quad (6)$$

Owing to the limited memory in FPGA hardware resources, measurements for each pattern have been calculated at simulation stage, but this fact does not invalidate experimental results.

In Fig. 12, values of SE with $n$ from 1 to 3 and for patterns A, B, C and D are shown. These values are compared with a fifth case named E, which corresponds to the randomized signal after encryption of pattern A, which can be considered the worst case in terms of entropy. In this last pattern E, it is possible to notice that SE is, as expected, almost the ideal value 8.0606. However as the transmission rate decreases from D to A, SE value also decrease. It is a logical result, as when lower bandwidth transmission is used, IFGs full of IDLES takes more bandwidth percentage versus the random payloads of the transmitted frames.

Thank to this results, it is possible to conclude that encryption works and makes indistinguishable data traffic patterns, which permits to hide the pattern and contents of Ethernet traffic from any passive eavesdropper.

## VII. Conclusion

As far as the authors are aware, this is the first time that an encryption mechanism is proposed for ciphering physical layer communications based in 8b/10b coding. This new encryption system consists of symmetric ciphering of the complete 8b/10b symbol stream. Encryption based on an FPE blockcipher working in CTR mode has been simulated and implemented. Also security considerations have been taken into account. The proposed system is able to obfuscate the data traffic pattern, which could improve the overall security, with no throughput losses, null overhead and low latency.

These properties make this kind of 'in-flight' encryption suitable for protocols where delay and jitter are critical, such as real time Ethernet in industrial environments, which would improve security in modern optical SCADA networks.

In addition to this, by preserving coding properties as DC-balance, short run length, and transition density, physical layer encryption is achieved easily without making changes in the subsequent hardware elements or medium dependent circuitry. As an example, in Ethernet 1000Base-X standard, SERDES (Serializer/Deserializer) and commercial SFP (Small Form-Factor Pluggable) optical modules could be compatible with the proposed encryption system.

Regarding to its implementation, although in the proposed system there is an increment in the hardware resources with respect a standard AES blockcipher, this system entails a ratio throughput/resources better than existing FPE implementations.

Finally, as future work, authors consider to study the implementation of physical layer encryption for other Ethernet standards with higher transmission rates such as 10G-BaseR.

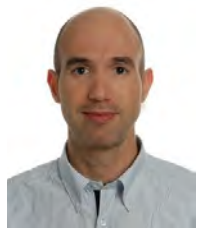
**Adrián Pérez-Resa** was born in San Sebastián, Spain. He received the M.Sc degree in Telecommunications Engineering from the University of Zaragoza, Zaragoza, Spain, in 2005.
He worked in Telecommunications industry as R&D engineer for more than ten years, nowadays he is a Ph.D candidate and a member of the Group of Electronic Design, Aragón Institute of Engineering Research, University of Zaragoza. His research interests include high speed communications and cryptography applications.

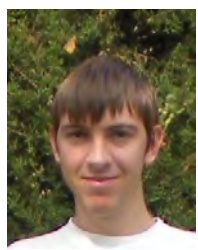
**Miguel Garcia-Bosque** was born in Zaragoza, Spain. He received the B.Sc and M.Sc in Physics from the University of Zaragoza, Zaragoza, Spain in 2014 and 2015 respectively.
He is currently a member of the Group of Electronic Design, Aragón Institute of Engineering Research, University of Zaragoza. His research interests include chaos theory and cryptography algorithms.

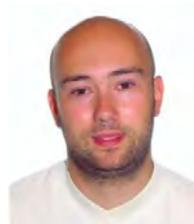
**Carlos Sánchez-Azqueta** was born in Zaragoza, Spain. He received the B.Sc., M.Sc., and Ph.D. degrees from the University of Zaragoza, Zaragoza, Spain, in 2006, 2010, and 2012, respectively, all in Physics, and the Dipl.-Ing. Degree in electronic engineering from the Complutense University of Madrid, Madrid, Spain in 2009.
He is currently a member of the Group of Electronic Design, Aragón Institute of Engineering Research, University of Zaragoza. His research interests include mixed-signal integrated circuits, high-frequency analog communications, and cryptography applications.

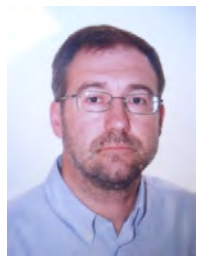
**Santiago Celma** was born in Zaragoza, Spain. He received the B.Sc., M.Sc., and Ph.D. degrees in physics from the University of Zaragoza, Zaragoza, Spain, in 1987, 1989, and 1993, respectively.
He is currently a Full Professor in the Group of Electronic Design, Aragon Institute of Engineering Research, University of Zaragoza. He has coauthored more than 100 technical papers and 300 international conference contributions. He is coauthor of four technical books and the holder of four patents. He appears as principal investigator in more than 30 national and international research projects. His research interests include circuit theory, mixed-signal integrated circuits, high-frequency communication circuits, wireless sensor networks and cryptography for secure communications.